\documentclass{iau}

\usepackage{amsmath}
\usepackage{graphicx}
\usepackage{multirow}
\usepackage{blindtext}
\begin{document}

\lefttitle{C. Saha \& D. Nandy}
\righttitle{Understanding Grand Minima in Solar Activity}

\jnlPage{1}{7}
\jnlDoiYr{2021}
\doival{10.1017/xxxxx}

\aopheadtitle{Proceedings IAU Symposium}
\editors{A. V. Getling \& L. L. Kitchatinov, eds.}

\title{Understanding Grand Minima in Solar Activity: Confronting Observations with Dynamo Simulations}

\author{Chitradeep Saha$^{1}$ and Dibyendu Nandy$^{1,2}$}
\affiliation{$^1$Center of Excellence in Space Sciences India, Indian Institute of Science Education and Research Kolkata, Mohanpur 741246, West Bengal, India}

\affiliation{$^2$Department of Physical Sciences, Indian Institute of Science Education and Research Kolkata, Mohanpur 741246, West Bengal, India}

\begin{abstract}

The grand minimum in the Sun’s activity is a distinctive mode characterized by a magnetic lull that almost completely lacks the emergence of sunspots on the solar surface for an extended duration. The factors driving this transition of an otherwise magnetically active star into a quiescent phase, the processes occurring within the solar interior and across the heliosphere during this period, and the mechanisms leading to the eventual resurgence of surface magnetic activity remain enigmatic. However, there have been sustained efforts in the past few decades to unravel these mysteries by employing a combination of observation, reconstruction and simulation of solar magnetic variability. Here, we summarize recent research on the solar grand minimum and highlight some outstanding challenges -- both intellectual and practical --  that necessitate further investigations.

\end{abstract}


\maketitle

\section{Introduction}

The Sun is a magnetized star that exhibits variability in its magnetic activity across various timescales ranging from minutes to millennia and beyond \citep{NANDY2021PEPS, PEVTSOV2023ADVSPRES, BISWAS2023SPSCREV}. The characteristic decadal variation in solar magnetic activity, i.e.,
the $\sim$11-yr Schwabe cycle is well observed and recorded for multiple centuries \citep{Hathaway2015LIVREV}. Besides, there have been several instances of intermittent periods characterized by a magnetically quiescent state of the Sun persisting over decades and even centuries -- known as the solar grand minimum -- as evidenced in reconstructed solar activity using indirect proxies such as cosmogenic isotopes and auroral activity records \citep{Solanki2004Nature, Steinhilber2009GRL, Usoskin2023LIVREV}. 

\begin{figure}[h!]
    \centering
    \includegraphics[width=\textwidth]{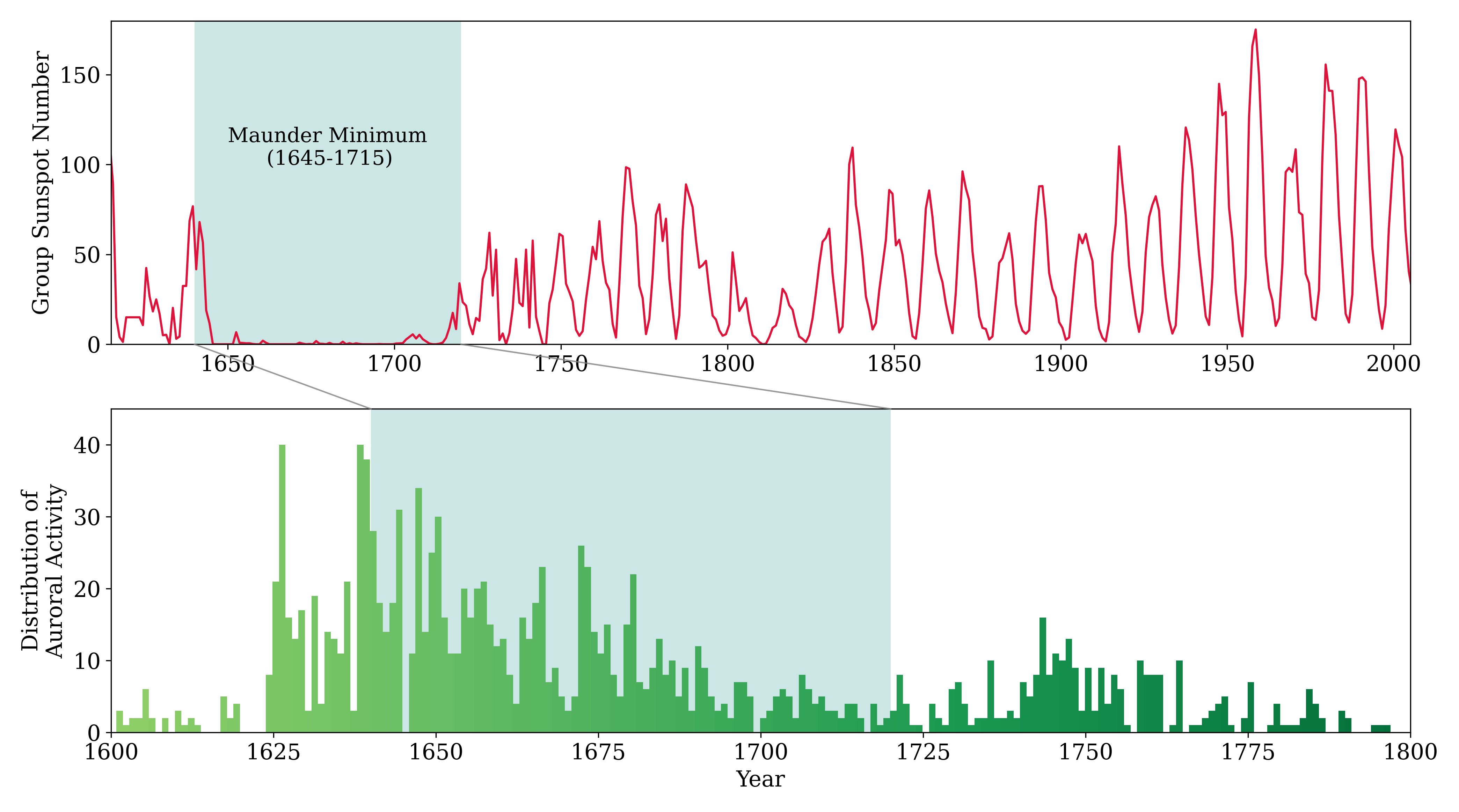}
    \caption{Top panel: group sunspot number time series (top panel) highlighting one of the most recent solar grand minima during 1645-1715 -- known as the Maunder minimum \citep{EDDY1983SOLARPHYSICS}. Bottom panel: Temporal distribution of auroral activity records \citep{WANG2021JGR} suggesting a persistent solar activity modulation during Maunder minimum.}
    \label{fig:01}
\end{figure}

From an observational perspective, solar grand minima are marked by a significant decrease in sunspot numbers (see, Fig.\ref{fig:01}) and a reduction of large-scale solar magnetic field strength and open flux for a prolonged period \citep{VAQUERO2015ADVSPRES, Carrasco2019ApJ, Usoskin2021AandA}. Historically, solar grand minima have been associated with periods of cooler temperatures on Earth, as evidenced by the temporal proximity of the Maunder Minimum (1645–1715) and the Little Ice Age, hinting towards a potential link between solar grand minima and terrestrial cooling \citep{EDDY1976SCIENCE}. However, whether the latter is causally concomitant with the former is debatable \citep{OWENS2017JSWSC}. Nevertheless, the Sun being the primary driver of heliospheric space environmental conditions, variabilities in solar open flux eventually influence the space weather and space climate over short and long timescales, respectively \citep{SCHRIJVER2015ADVANCESINSPACERESEARCH, NANDY2023JASTP}.

On the other hand, the precision in modeling regular solar-like activity cycles and predicting the amplitude and progression of upcoming cycles has significantly advanced in recent decades \citep{BHOWMIK2018NATCOM, PETROVAY2010LIVREV, NANDY2021SOLPHYS}. In contrast, the ability to predict the onset of solar grand minimum and its duration has yet to achieve substantial success. In fact, the unusually long minimum of sunspot cycle 23 \citep{NANDY2011NATURE} prompted debate and speculation among researchers about the potential occurrence of an impending solar grand minimum \citep{SOLANKI2011SCIENCE, ZOLOTOVA2014JGR}.  

It is now widely acknowledged that there operates a dynamo mechanism in the solar convection zone (SCZ), and it is responsible for the periodic generation and recycling of large-scale solar magnetic fields \citep{Charbonneau2020Livrev}. Owing to the extreme physical conditions in the SCZ, such as density and temperature stratification, turbulent plasma motions, magnetic stresses and back-reactions, among others, the dynamo mechanism exhibits irregular behavior. This irregularity is believed to account for the observed variability in the strength and timing of sunspot cycles.  As a theoretical limit, it is hypothesized that this irregularity may occasionally intensify to the extent that it can drive the dynamo operation below a critical threshold, thereby causing an intermittent behavior in the solar magnetic activity -- the solar grand minimum \citep{PASSOS2014AAP}. A more detailed discussion on various approaches towards modeling solar activity extrema follows in section 3.

\begin{figure}[h!]
    \centering
    \includegraphics[width=\textwidth]{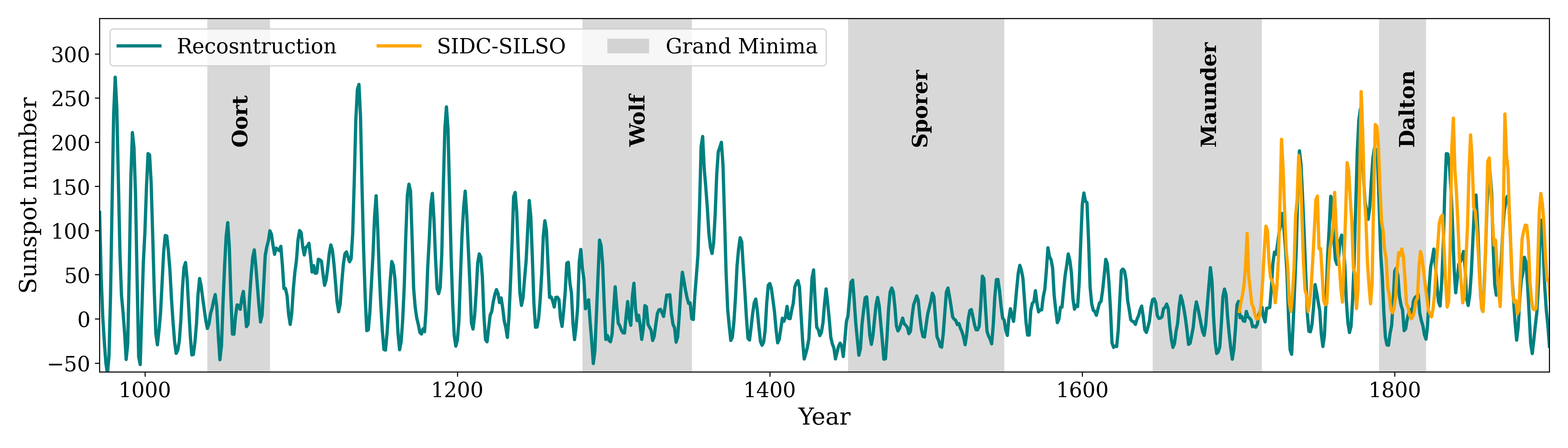}
    \caption{Direct sunspot records complemented by solar activity reconstructions using $^{14}$C cosmogenic isotope put forward evidence of multiple solar grand minima (shaded in gray) that have occurred over the past millennium \citep{Usoskin2021Vizier}.}
    \label{fig:03}
\end{figure}

The primary challenge in understanding solar grand minima is posed due to insufficient observational constraint on dynamo models, unlike regular solar activity cycles \citep{Jaramillo2019NatAS}. Thanks to various terrestrial archives -- such as tree rings and ice cores -- that preserve the millennial timescale solar activity signatures and give significant insights into solar magnetic behavior during grand minima (see, Fig.\ref{fig:01}\&\ref{fig:03}). However, numerical simulations remain indispensable for gaining knowledge of the dynamics during such phases in the Sun’s uncharted territories, such as its interior and polar regions.

In the subsequent sections, we aim to revisit our current understanding of the solar grand minimum in terms of both numerical simulations and observations, emphasize the complementarity between reconstruction, observation and physics-based numerical simulations of extreme solar activity, discuss critical insights gleaned from recent results and their implications on our space environment. We also list down some of the outstanding aspects of solar/stellar grand minima that are required to be explored further.

\section{Solar grand minimum: Fact or Artifact? }

Before investigating deeper into solar grand minima, a common skepticism arises whether these episodes of reduced sunspot numbers genuinely exist or they are merely artifacts resulting from limited and sparse observational data points \citep{HOYT1996SOLPHYS}. Based on historical records, \cite{Zolotova2015ApJ} argued that the Maunder Minimum could be an ordinary secular minimum with diminished decadal variability. \citep{Feynman2011SolPhys} claimed the Maunder Minimum to be the trough of the centennial Gleissberg cycle. Even if grand minima do exist, there remains considerable uncertainty and a lack of consensus regarding their impact on terrestrial cooling and climatology \citep{OWENS2017JSWSC}.

Nevertheless, sophisticated methods of solar magnetic activity reconstruction based on various proxies like the terrestrial abundance of cosmogenic isotopes reveal that there were indeed multiple episodes of critically low magnetic flux output from the Sun for prolonged duration over the past few millennia \citep{Steinhilber2009GRL, USOSKIN2015AandA, Usoskin2023LIVREV}. Data gleaned from these sources could be calibrated well with the modern-day sunspot record, suggesting the reliability of the reconstruction techniques. Recently, \cite{HAYAKAWA2024MNRAS} explored ancient sunspot records at the onset of the Maunder minimum that corroborate well with the proxy records. Evidence of declining auroral activity during the deep phase of the Maunder Minimum put independent support to the reduced solar activity during this time \citep{WANG2021JGR} (see, Fig.\ref{fig:01}).

Moreover, multiple works have theorized the solar grand minimum as an outcome of the irregularity in the dynamo mechanism and have successfully reproduced many of the observational signatures and statistics of solar grand minimum \citep{PASSOS2014AAP, TRIPATHI2021MNRASL, SAHA2022MNRASL}. This strengthens the scenario in favour of the solar grand minimum being a reality and prompts that there may be possibilities of impending intermittency again in the solar variability.

\cite{Jaramillo2019NatAS} presented an exhaustive summary of various observational datasets covering the past few centuries  and their collective reliability. While it is a fact that there exist not many coherent observations of the Sun during the Maunder Minimum phase, a period of prolonged quiescence emerges in all the datasets, suggesting that one cannot completely rule out the occurrence of solar grand minima \cite{USOSKIN2015AandA}.

\section{Numerical modeling of intermittent solar activity: Current understanding}

In a cursory view, solar grand minima episodes are mainly devoid of surface eruptions of sunspots. This phenomenon can theoretically be modelled as a reduced abundance of bipolar magnetic sources in the solar surface flux transport models \citep{MACKAY2003SOLPHYS, WANG2003APJ}. Such models regulate the production rate of magnetic bi-poles at different phases of sunspot cycles to mimic observed statistics of the depth and duration of grand minima \citep{USOSKIN2007AANDA}. However, the underlying reasons for such non-uniform regulation are not imbibed into these models.

One of the proposed avenues to numerically model grand minima like extreme solar variability is to introduce self-consistent irregularities, i.e., stochastic forcing on the dynamo processes \citep{MOSS2008SOLPHYS, USOSKIN2009SOLPHYS}. These irregularities predominantly manifest during the induction of large-scale poloidal fields. Therefore, fluctuation in poloidal source intensity across the solar hemispheres can drive the solar dynamo into a sub-critical regime producing intermittent grand minimum-like episodes \citep{CHARBONNEAU2004APJ, OLEMSKOY2013AREP, BRANDENBURG2008ANOTES, PASSOS2014AAP, SAHA2022MNRASL}. These models incorporate random fluctuations in the poloidal source, in the Babcock-Leighton framework or the electromotive force in the Parker-type mean-field dynamo framework as a signature of stochastic forcing. Spatially reduced simpler time-delay dynamo models can also simulate the entry and recovery from grand minima episodes when subjected to stochastic driving \citep{WILMOTSMITH2005MNRAS, WILMOTSMITH2006APJ,  HAZRA2014APJ, TRIPATHI2021MNRASL}.

Magnetically buoyant flux tubes produced deep in the SCZ rise upward and are subject to turbulent buffeting by the convective plasma flows. This eventually introduces a dispersion around the mean tilt angle distribution of sunspots \citep{DASIESPUIGAANDA2010, NAGY2017SOLPHYS, Pal2023APJ}. Migration and diffusion of randomly tilted bipolar magnetic regions on the solar surface eventually influence the polar field production rate and, in turn, the strength of the next sunspot cycle \citep{YEATES2008APJ}. Large anomaly in the bipoles can potentially simulate grand minima episodes \citep{KARAK2018APJL}. 

In axisymmetric flux transport dynamo models, a reduction in meridional circulation flow speed either in seclusion or in conjunction with fluctuating poloidal sources can replicate grand minimum-like activity phases \citep{Karak2010ApJ, CHOUDHURI2012PRL}. However, the observed variation in this weak flow is relatively small. More importantly, due to active region inflows, the meridional circulation is observed to be faster during solar activity minimum and vice-versa as shown by \cite{Hathaway2010Science}. They also pointed out a fundamental limitation of flux transport dynamo models in order to explain this discrepancy. Some more recent observational evidence indicating accelerated dynamics in solar activity during grand minima \citep{YAN2023AGUADVANCES, VELASCOHERRERA2024ADVSPRES} challenges the proposition of weak meridional flow as a cause of this intermittency.

The nonlinear back-reaction of magnetic fields on plasma flows in the convection zone is a plausible mechanism for generating magnetically quiescent phases \citep{Inceoglu2017ApJ}. Mathematically, this phenomenon can be modeled by incorporating either Lorentz feedback, algebraic $\alpha$-quenching, $\Lambda$-quenching, or a combination of these effects  \citep{TOBIAS1996AAP, KUKER1999AANDA, OSSENDRIJVER2000AANDA, SIMARD2020JSWSC}. However, the solar dynamo is shown to be only weakly nonlinear. Additionally, the random fluctuations are shown to play a more crucial role in explaining long-term solar variabilities \citep{CAMERON2019AANDA}.

Only a few studies have been conducted to investigate extreme solar activity using MHD simulation framework. \cite{Augustson2015ApJ} demonstrated episodes resembling grand minima in a convection-driven stellar dynamo model. Achieving a self-consistent onset and recovery of a Sun-like star into a grand minimum has yet to be accomplished in direct numerical simulations -- underscoring that our theoretical comprehension of grand minima and the overall solar dynamo mechanism is not yet complete.

\section{Persistence of dynamo activity during solar grand minimum: Observational and numerical evidence}

Does the underlying dynamo mechanism cease when the solar surface magnetic activity slips to a slumber?  For the first time, a reasonable clue was put forward by \cite{BEER1998SOLPHYS} in the form of observational evidence suggesting a persistent decadal timescale modulation in reconstructed solar activity, even during grand minima. Later, several other reconstructions have corroborated this finding \citep{MIYAHARA2004SOLPHYS, USOSKIN2015AandA,  Usoskin2023LIVREV}.

\begin{figure}[h!]
    \centering
    \includegraphics[width=\textwidth]{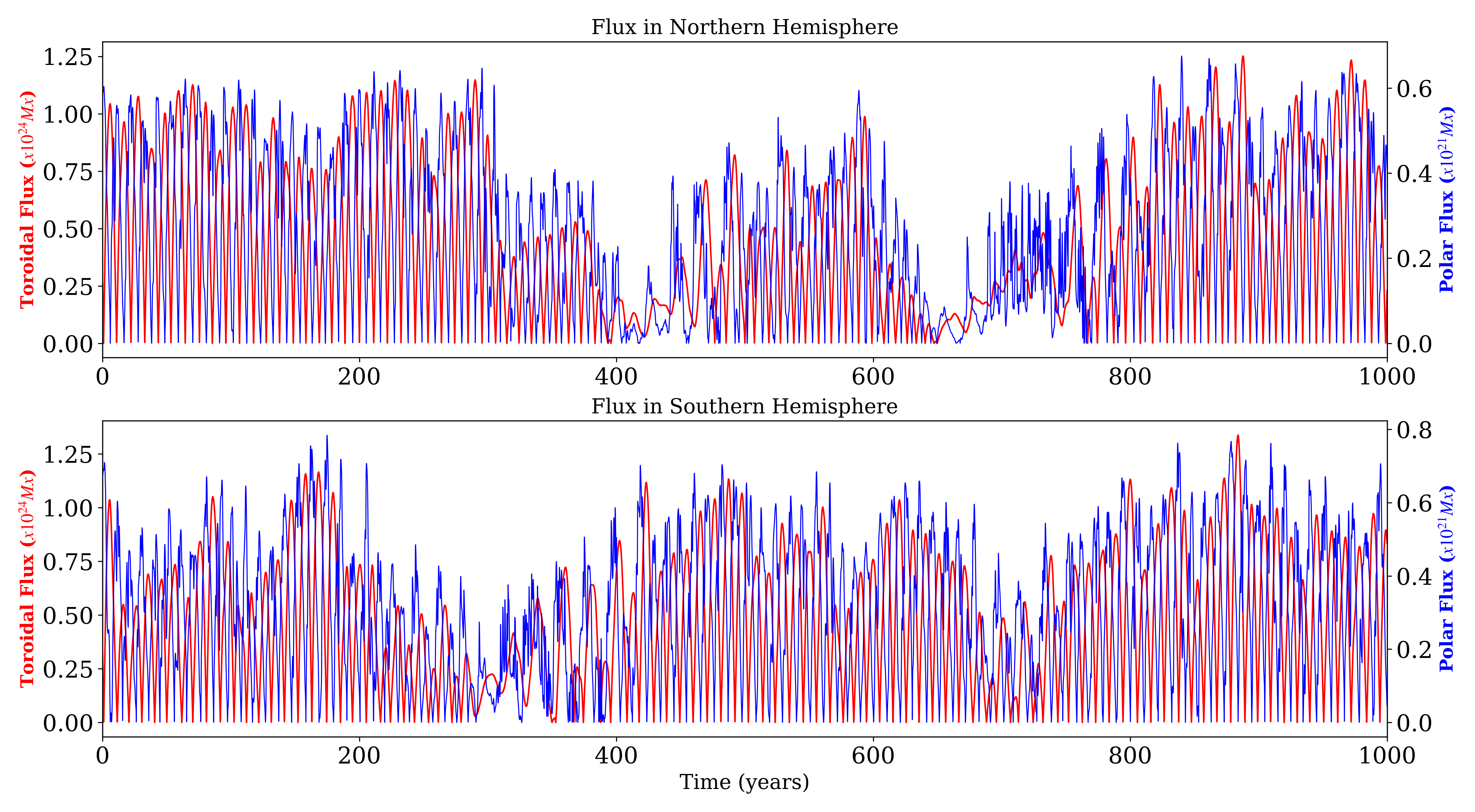}
    \caption{Results from stochastic numerical dynamo simulation by \cite{SAHA2022MNRASL} showing the evolution of solar poloidal and toroidal magnetic flux. Multiple episodes of grand minima with significantly low magnetic output are captured independently in the solar northern (top panel) and southern (bottom panel) hemispheres.}
    \label{fig:04}
\end{figure}

Numerical simulations like \cite{CHARBONNEAU2004APJ, SAHA2022MNRASL} have shown persistence in phase and cyclicity during intermittent episodes (see, Fig.\ref{fig:04}).
An intriguing aspect of simulating this persistence of magnetic activity and eventual recovery from a solar grand minimum in flux transport dynamo models is to introduce an additional polar field generation mechanism slightly deeper in the convection zone in addition to the Babcock-Leighton source term near the solar surface \citep{PASSOS2014AAP, SAHA2022MNRASL}.  This is to compensate for the deficit of polar flux near the solar surface due to the absence of sunspots therein. A downward magnetic pumping in the solar convection zone also can help recover the Sun from grand minima \citep{KARAK2018APJL}.

 However, the 11-yr sunspot cycle disrupts to some extent, and power distribution in different periodicities considerably re-organizes during grand minima. Several shorter and longer periodic phenomena become prominent, as elucidated by  \cite{SAHA2022MNRASL}. We will elaborately discuss this in the upcoming sections.

\section{Hemispheric asymmetry in solar activity during grand minimum}

\begin{figure}[h!]
    \centering
    \includegraphics[width=\textwidth]{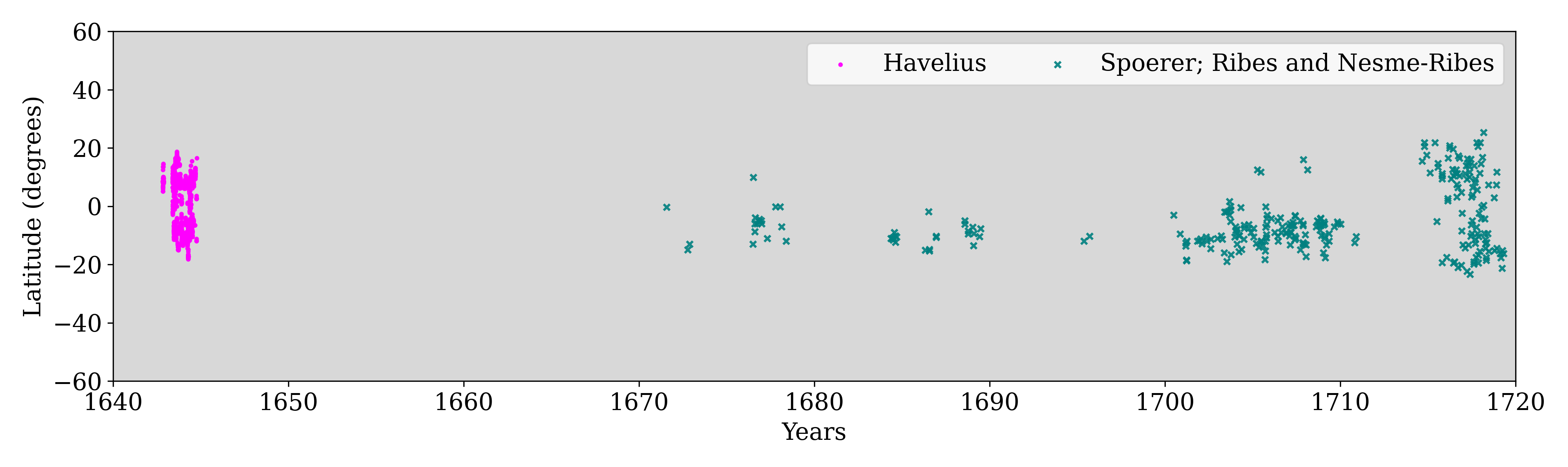}
    \caption{Sunspot butterfly diagram constructed from historical observational datasets \citep{RIBES1993AAP, VAQUERO2015ADVSPRES, Carrasco2019ApJ}, depicting strong hemispheric asymmetry in the solar activity during Maunder minimum with a prolonged preferential bias in the latitudes of sunspot emergence -- predominantly confined to the southern solar hemisphere. }
    \label{fig:07}
\end{figure}

Evidence from helioseismic measurements hints towards a nearly zero meridional plasma flow at the solar equator, putting an important constraint on solar dynamo models \citep{JARAMILLO2009APJ}. This implies that the solar hemispheres are weakly coupled, predominant contribution of which comes from the cross-equatorial diffusion of magnetic fields over a longer timescale. Moreover, the meridional circulation itself demonstrates a hemispheric asymmetry as it evolves, plausibly causing an asymmetry in solar activity across the two hemispheres \citep{LEKSHMI2018APJ}. Such hemispheric asymmetry causes a deviation from the dipolar parity in large-scale solar magnetic fields. This effect is more pronounced during solar grand minima as confirmed by both simulations \citep{HAZRA2019MNRAS} and observations (see, Fig. \ref{fig:07}).

\cite{OLEMSKOY2013APJ} simulated asymmetric solar activity during a grand minimum by randomizing the preferred latitude of sunspot emergence in the active latitudinal belts across the hemispheres. Another justifiable approach is to introduce stochastic fluctuations independently into the poloidal sources in the two hemispheres, resulting in the asymmetric occurrence of hemispheric grand minima  \citep{PASSOS2014AAP, SAHA2022MNRASL}.

\section{Probing the solar interior and polar dynamics}

\begin{figure}[h!]
    \centering
    \includegraphics[width=\textwidth]{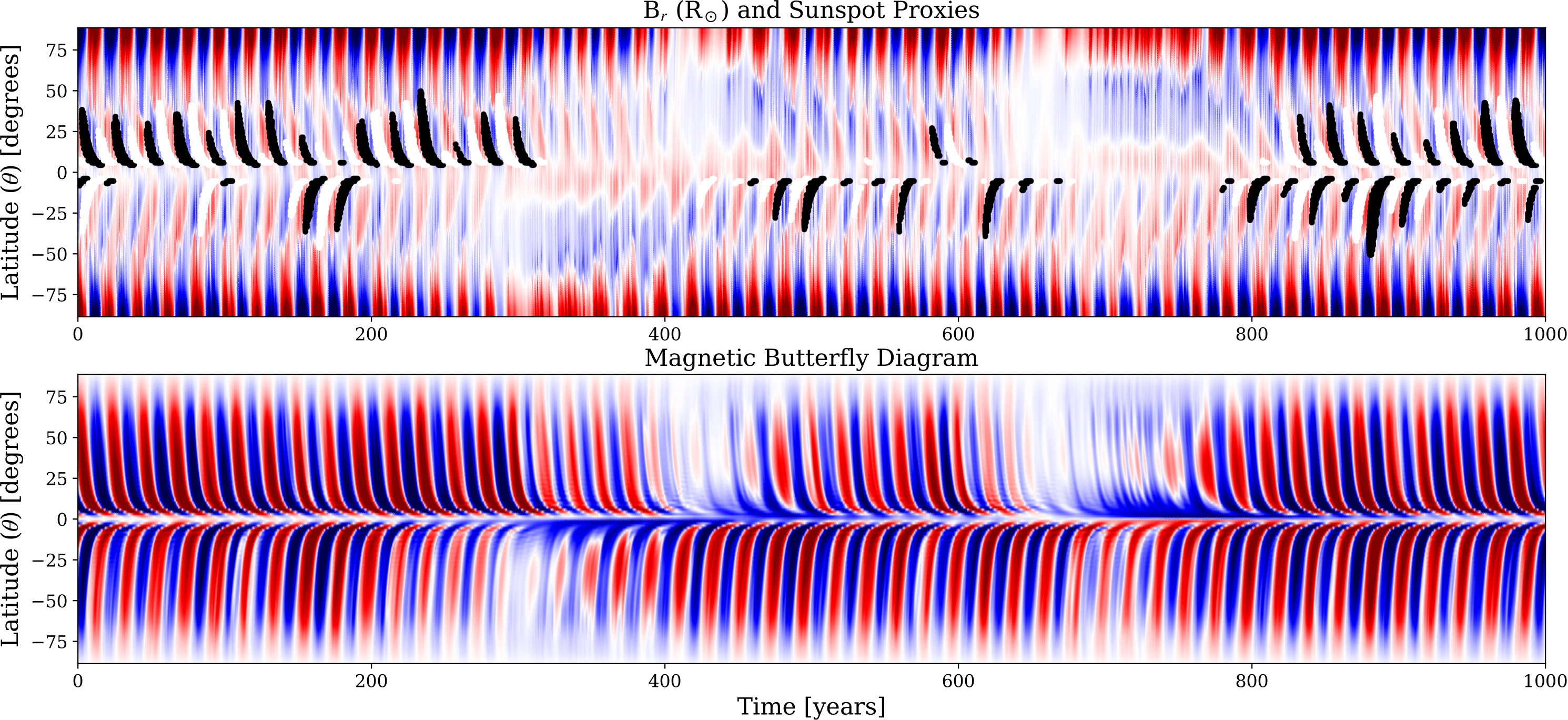}
    \caption{Stochastically forced numerical solar dynamo simulations can potentially provide novel insights into the dynamics on the Sun's surface including the polar regions (top panel) and at the base of the convection zone (bottom panel) during regular and extreme activity phases, direct observations of which are not available otherwise (ref. \citealp[]{SAHA2022MNRASL}).}
    \label{fig:05}
\end{figure}

Direct probing of the solar interior is not possible. On the other hand, existing high latitude observations of the Sun also suffer from large projection effects \citep[and references therein]{NANDY2023BAAS}. These make the solar interior and polar regions almost uncharted territories. Knowledge about the solar interior and poles is crucial because, in the absence of sunspots in the active latitudes during grand minima, the polar and deep-seated toroidal magnetic fields determine the dynamics of solar activity.  Numerical simulations of solar dynamo are essential in this context. Multiple solar dynamo simulations have independently reported instances of a temporary halt in the polar field reversal during deep solar grand minima \citep{MACKAY2003SOLPHYS, SAHA2022MNRASL} (see, Fig.\ref{fig:05}).   The reversal can eventually resume due to gradual accumulation of magnetic fluxes from ephemeral regions and pores \citep{VSVANDA2016AANDA} transported by the poleward branch of meridional circulation \citep{SAHA2022MNRASL}, until sufficient polar magnetic flux is built up to kick start regular sunspot cycles.

\section{Grand minima through a mathematical prism: Spectral components of solar grand minima}

\begin{figure}[h!]
    \centering
    \includegraphics[width=\textwidth]{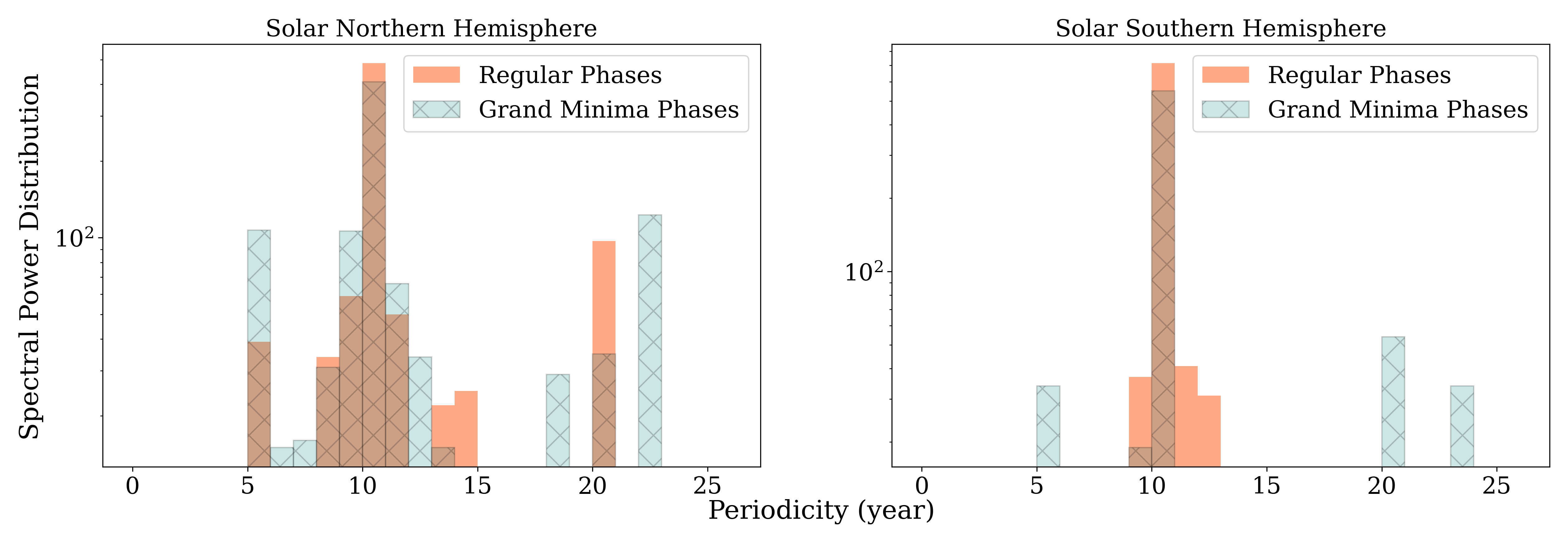}
    \caption{Spectral power distribution indicates towards a relative suppression of 11-year cyclicity in solar activity and enhancement of other shorter and longer periodicities during simulated solar grand minima. see, \cite{SAHA2022MNRASL} for further details.}
    \label{fig:06}
\end{figure}

Spectral analysis techniques, when applied to reconstructed and simulated solar open flux time series, provide valuable insights into the dominant periodicities in solar magnetic activity, including that of during grand minima \citep{MIYAHARA2004SOLPHYS, INCEOGLU2015AANDA, Usoskin2021AandA}. While the spectral power stored in the 11-year periodicity diminishes during grand minima, several other shorter and longer periodicities are enhanced \citep{SAHA2022MNRASL}. Recent studies on auroral activity records indicate towards a shortened solar cycle period during the Maunder minimum \citep{YAN2023AGUADVANCES, VELASCOHERRERA2024ADVSPRES}. This can be explained by faster meridional flows during grand minima, resulting in an increased rate of magnetic flux transportation, as discussed in section 3. 

Previous studies showed that a deep meridional flow acts as a clock and regulates the solar cycle timescales \citep{NANDY2002SCIENCE, HATHAWAY2003APJ}. Recently, \cite{SAHA2022MNRASL} speculated the ceaseless meridional plasma flows in tandem with mean-field $\alpha$-effect to be responsible for the dynamo operation to recuperate from a grand minimum. In fact, the signature of meridional plasma motion is captured in the spectral domain in the form of a $\sim$5-yr component (see, Fig. \ref{fig:06}), which is also the characteristic timescale to dredge up magnetic fields from the tachocline to the active latitude on the solar surface solely by the meridional circulation in the absence of any buoyancy mechanism \citep{SAHA2022MNRASL}. Interestingly, a subsequent study by \cite{INCEOGLU2024SCIREP} has discovered a robust presence of a similar spectral component in a reconstructed solar activity dataset with high temporal resolution.

\section{Influence of solar grand minima on the space environmental conditions}

\begin{figure}[h!]
    \centering
    \includegraphics[width=0.9\textwidth]{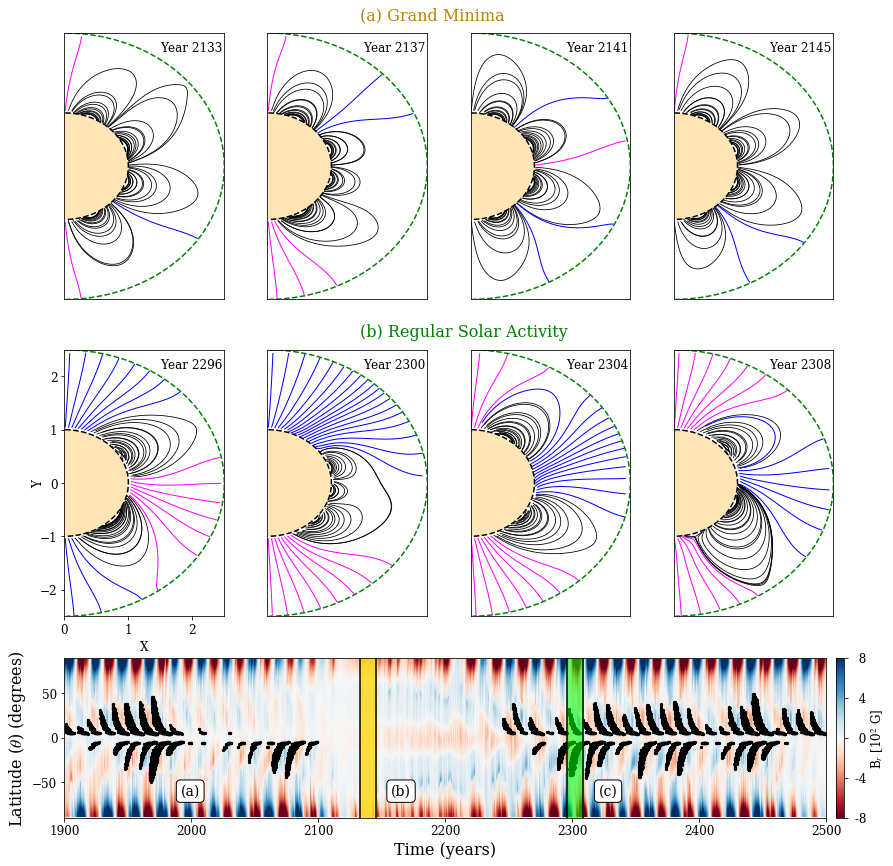}
    \caption{Evolution of large-scale magnetic field configuration in the solar corona during different levels of solar activity including grand minimum (ref. \citealp[]{DASH2023MNRAS}).}
    \label{fig:08}
\end{figure}

The large-scale magnetic field in the solar corona acts as a bridge between the dynamo-generated fields in the solar interior and the interplanetary magnetic fields pervading all through the heliosphere, thereby establishing a causal connection between the solar interior and the state of the heliosphere, as elucidated by \cite{NANDY2023JASTP}. Therefore, it goes without saying that a solar grand minimum can profoundly influence the space environmental conditions \citep{OWENS2012GRL, Riley2015ApJ, Hayakawa2021JSWSC}.

With the aid of potential field source surface extrapolations \cite{DASH2023MNRAS} have illuminated that during these periods of reduced solar magnetic activity, the heliosphere may experience significant changes -- the weakening of the solar open magnetic flux, leading to decreased solar wind pressure and interplanetary magnetic field strength. This can alter the Earth's magnetosphere, ionosphere, and thermosphere structure and dynamics. Furthermore, the reduced solar activity during grand minima is associated with decreased occurrence and intensity of solar flares and coronal mass ejections \citep{Richardson2012JSWSC}. 

The topology of the large-scale solar coronal magnetic field during a grand minimum (see, Fig.\ref{fig:08}, top panel) can be strikingly different from that of the regular solar cycle activity minimum (see, Fig.\ref{fig:08}, middle panel, Year 2296 and 2308). The closed magnetic loop-like structures in the solar corona predominantly manifest throughout the heliographic latitudes, including the polar regions, unlike a dipolar configuration during solar minimum. This reasonably explains the unusually large number of auroral events during the Maunder minimum \citep{WANG2021JGR}.

\section{Outstanding questions and concluding remarks}

There are still multiple unresolved inquiries about solar grand minima, leading to continued research endeavours to enhance our comprehension of these events. A few of these inquiries consist of: 

    1. What are the primary mechanisms responsible for initiating grand minima events in the solar cycle? Understanding the triggers behind these prolonged periods of reduced solar activity is essential for predicting their occurrence and assessing their potential impacts.

    2. How do solar dynamo processes behave during grand minima events? Investigating the behavior of the solar dynamo during periods of reduced magnetic activity can provide insights into the underlying mechanisms governing the solar cycle.
    
    3. What factors determine the duration and frequency of grand minima events? Investigating the variability in the length and recurrence of grand minima can provide insights into the underlying processes driving solar cycle dynamics.

    4. What is the relationship between solar grand minima and terrestrial climate variability? Understanding the linkages between prolonged periods of reduced solar activity and climate changes on Earth can shed light on the mechanisms driving long-term climate trends.

Addressing these outstanding questions is crucial for advancing our understanding of solar grand minima and their implications for space weather, climate variability, and solar dynamo processes.

\section{Acknowledgements}

C.S. is grateful to have received a travel grant from the International Astronomical Union to attend the IAU Symposium 365 in Yerevan, Armenia. C.S. also acknowledges the financial support from CSIR through grant no. 09/921(0334)/2020-EMR-I. CESSI is funded by IISER Kolkata, Ministry of Education, Government of India.

\bibliographystyle{iaulike}
\bibliography{references}

\end{document}